\documentclass[conference]{IEEEtran}
\usepackage{cite}
\usepackage{amsmath,amssymb,amsfonts}
\usepackage{algorithmic}
\usepackage{graphicx}
\usepackage{textcomp}
\usepackage{xcolor}
\usepackage{url}
\usepackage{listings}
\graphicspath{{figs/}}
\usepackage[labelfont=bf,textfont=bf]{caption}

\def\BibTeX{{\rm B\kern-.05em{\sc i\kern-.025em b}\kern-.08em
    T\kern-.1667em\lower.7ex\hbox{E}\kern-.125emX}}
    
\usepackage{blindtext}

\begin{document}

\title{Adding Explicit Load-Acquire and Store-Release Instructions to the RISC-V ISA}

\author{\IEEEauthorblockN{Bryce Arden\IEEEauthorrefmark{1}, Zachary Susskind\IEEEauthorrefmark{1}, Brendan Sweeney\IEEEauthorrefmark{1}}
\IEEEauthorblockA{\textit{Electrical and Computer Engineering} \\
\textit{The University of Texas at Austin}\\
Austin, TX, USA \\
\texttt{\{arden.bryce,zsusskind,turtwig\}@utexas.edu}}
}

\maketitle
\begingroup\renewcommand\thefootnote{\IEEEauthorrefmark{1}}
\footnotetext{Authors contributed equally}
\endgroup


\begin{abstract}
Weak memory models allow for simplified hardware and increased performance in the memory hierarchy at the cost of increased software complexity. In weak memory models, explicit synchronization is needed to enforce ordering between different processors. Acquire and release semantics provide a powerful primitive for expressing only the ordering required for correctness. In this project, we explore adding \textit{load-acquire} and \textit{store-release} instructions to the RISC-V ISA. We add support to the \texttt{herd} formal memory model, the gem5 cycle-approximate simulator, and the LLVM/Clang toolchain. Because these instructions do not exist in the RISC-V standard, there is an inherent urgency to ratify explicit \textit{load-acquire/store-release} instructions in order to prevent multiple ABI implementations and ecosystem fragmentation. We found that for workloads with a high degree of sharing and heavy contention, the impact of less memory ordering is muted, but our changes successfully encode the semantics we desire.

\end{abstract}

%
%

\section{Introduction}
\label{sec:introduction}
RISC-V~\cite{Waterman2016DesignOT} is a free, open-source Instruction Set Architecture (ISA) which is growing in popularity across all application domains, ranging from datacenters, to clients, to embedded devices.
RISC-V defines and uses a memory consistency model called RVWMO (RISC-V Weak Memory Ordering)~\cite{riscv_manual}, which is weaker than those present in some other ISAs like x86, while stronger than those in POWER and ARM.
Under RVWMO, the memory operations of a hardware thread (or ``hart'') appear in-order from the perspective of that same hart, but may appear out-of-order from the perspective of other harts.
This creates challenges for writing functionally correct and performant code.

RISC-V is constantly evolving thanks to a robust extension framework~\cite{riscv_extensions}, which allows it to be adapted to target new software applications.
However, like any open-source project, RISC-V is susceptible to ecosystem fragmentation. Left unchecked, this could result in two processors, both of which are nominally using the RISC-V ISA, actually supporting radically different sets of instructions, which would undoubtedly harm software compatibility and hardware adoption.
As a consequence, RISC-V specifies a small set of \textit{profiles}~\cite{riscv_profiles}, which are intended to represent and standardize common design decisions for different domains.
Specifically, each profile is composed of the base RISC-V ISA, a set of mandatory extensions, and a small set of optional features.
While this approach reduces fragmentation, it does not entirely eliminate the issue of application binary interface (ABI) incompatibility, or any other potential complications which may be introduced during backend codegen. ABI breakage remains a challenge in the ecosystem, as a breaking ABI change requires downstream applications to be re-compiled.

In this project, we explore the addition of explicit \textit{load-acquire} and \textit{store-release} instructions as an extension to the RISC-V ISA. These instructions combine a memory operation with a \textit{one-way} memory barrier, enabling synchronization in a weak memory model while incurring less overhead than the traditional approach of using full or even partial bi-directional fences for synchronization. The ideas behind \textit{load-acquire} and \textit{store-release} instructions for weakly ordered memory consistency models are not new.
However, we believe that the absence of these instructions in RISC-V has negative implications for future compatibility at the ABI boundary, and this deficiency is worth addressing sooner rather than later for systems that choose to implement RVWMO.
The current upstream lowering for memory-consistent constructs produces unnecessarily restrictive fence instructions.
If we continue to ignore this issue, we will be encouraging individual vendors to provide their own ``more-performant'' ABIs, which introduces fragmentation and vendor lock-in.
Therefore, we see an unusual urgency for \texttt{ld.aq} and \texttt{st.rl} instructions in the RISC-V ISA.

Our specific contributions in this project are as follows:
\begin{enumerate}
    \item The augmentation of the RISC-V formal memory model with support for \texttt{ld.aq} and \texttt{st.rl}. This allows for a rigorous definition of what behaviors are allowed or forbidden in the presence of these instructions, which is of use for designing and verifying software and hardware.
    \item A proxy study to understand the performance impacts of load-atomic and store-release using Apple M1 silicon (AArch64). We use a modified version of the Clang C++ compiler which makes conservative memory ordering decisions and avoids the ARM-equivalent \texttt{LDAR} and \texttt{STLR} instructions.
    \item A study of the feasibility of using gem5 to simulate an extension to the RISC-V ISA incorporating \texttt{ld.aq} and \texttt{st.rl}. We find that although gem5 ``supports'' acquire and release semantics in AArch64 and RISC-V atomic memory operations, in practice, this support is non-performant due to limitations of the simulator memory model. It appears that significant changes to the gem5 weak memory model would be needed to simulate these operations with reasonably accurate performance.
\end{enumerate}

The remainder of this document is organized as follows: In Section \ref{sec:background}, we provide additional background on weak memory models, \textit{load-acquire} and \textit{store-release}, and prior work in this domain. In Section \ref{sec:methodology}, we discuss our experimental methodology for this project. In Section \ref{sec:results}, we present the augmented memory model, the results of the modified Clang compiler for AArch64, and our analysis of gem5. Lastly, in Sections \ref{sec:future_work} and \ref{sec:conclusion}, we discuss future work and conclude.
\section{Background and Related Work}
\label{sec:background}

The behavior of a parallel computer with a weak memory model is in some ways similar to that of a distributed system. Although a hart perceives its own local loads and stores as occurring in program order, it can not make assumptions about the ordering of events on other harts without explicit synchronization. A useful way to conceptualize memory accesses in this environment is as a partially ordered set, or \textit{poset}, ordered under Lamport's~\cite{lamport1978time} happened-before relation ``$\rightarrow$'', where for two events (e.g. memory accesses), $a$ and $b$:
\begin{enumerate}
    \item If $a$ and $b$ are events in the same hart, and $a$ occurred before $b$, $a \rightarrow b$.
    \item If $a$ and $b$ are events on separate harts, $a$ is a message send, and $b$ is the corresponding message receive, then $a \rightarrow b$.
    \item If $a \rightarrow b$ and $b \rightarrow c$, then $a \rightarrow c$.
\end{enumerate}
In a distributed system, remote events never become visible without an explicit message, which creates a happened-before relation. However, in a weak-memory system with shared memory, events from other harts will \textit{eventually} become visible even in the absence of synchronization, but without any deterministic ordering. Therefore, while the happened-before relation may seem conceptually simple, it can hide surprising complexities in the context of a weak memory model.

\begin{figure}[h]
 \includegraphics[width=1.18\columnwidth]{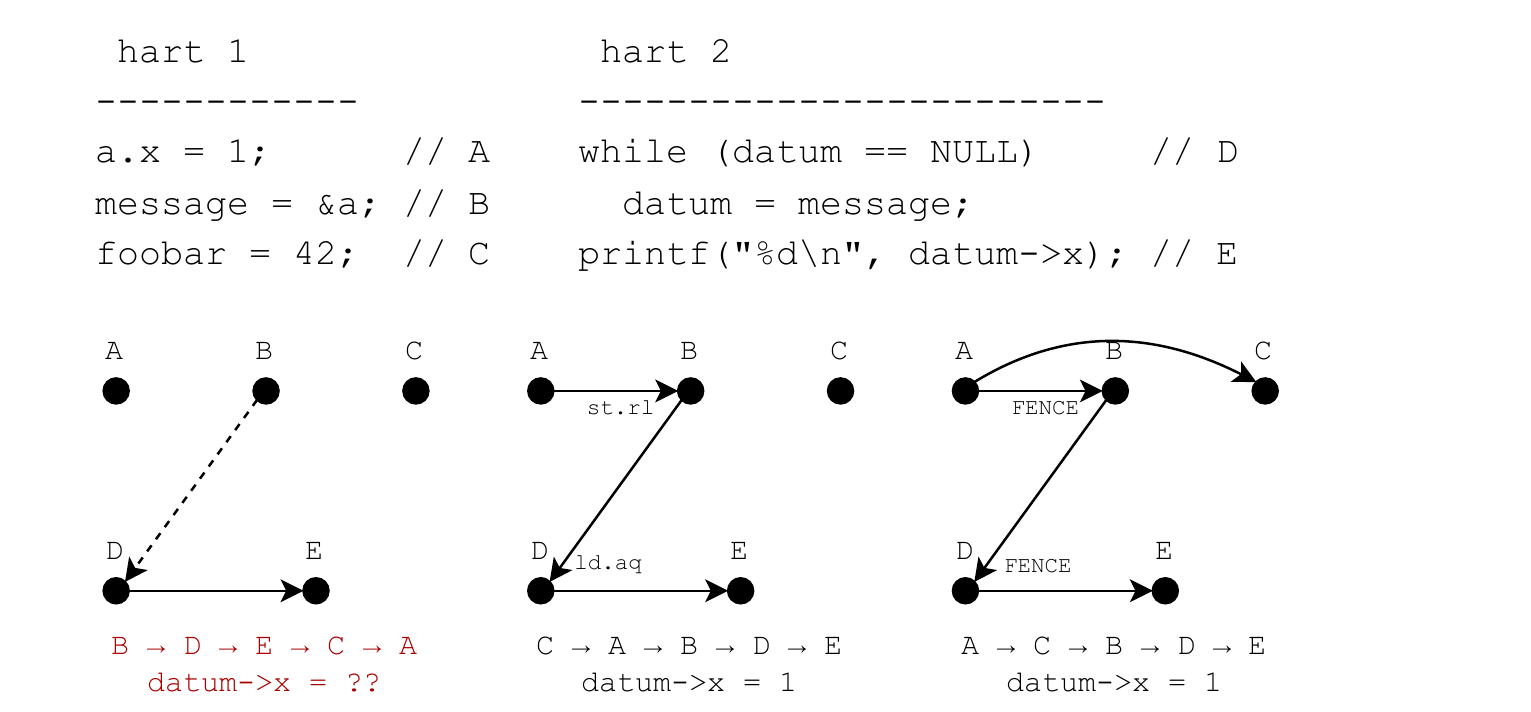}
 \caption{Possible event orders observed by hart 2 in the absence of synchronization, with synchronization via \texttt{ld.aq} and \texttt{st.rl}, and with synchronization via traditional memory fences. Code adapted from~\cite{bonzini_2021}.}
 \label{fig:synchronization}
\end{figure}

Consider the example shown in Figure \ref{fig:synchronization}. Even in the absence of explicit synchronization, hart 2 must observe event \texttt{B} (setting \texttt{message}) before event \texttt{D} (getting a non-null message). However, without further constraints, \texttt{E} may occur before \texttt{A} is observed, causing incorrect behavior. If a memory barrier is inserted in hart 1 between \texttt{A} and \texttt{B}, and in hart 2 between \texttt{D} and \texttt{E} (to prevent microarchitectural reordering of \texttt{E}), then the program behavior is functionally correct. However, the bi-directional barrier in hart 1 also prevents event \texttt{C} from being reordered before \texttt{A}, which would not impact functional correctness, and may have performance implications.

Using a store-release for \texttt{B} introduces a unidirectional fence which prevents prior events on hart 1 from being reordered after the \texttt{st.rl}. Using a \textit{load-acquire} for \texttt{D} introduces an opposite unidirectional fence, preventing subsequent events on hart 2 from being reordered prior to the \texttt{ld.aq}. Like the bidirectional fence-based solution, this is sufficient for functional correctness, but it imposes fewer constraints on the reordering of memory operations. Note that both approaches create an explicit happened-before relation between \texttt{B} and \texttt{D}, which is absent in the version without any synchronization. 

Arm included \textit{load-acquire} and \textit{store-release} semantics in the Armv8-A AArch64 ISA via the \texttt{LDAR} and \texttt{STLR} instructions~\cite{arm_docs}.
These instructions enable microarchitectural optimizations that are not possible with full barriers, reducing the performance impact of synchronization.

RISC-V already uses acquire (AQ) and release (RL) flags as part of the Atomic Memory Operation (AMO) instructions introduced in the RV32A and RV64A extensions~\cite{riscv_manual}. The ability to build on existing semantics further strengthens the case for a \textit{load-acquire/store-release} extension: since support for unidirectional barriers is already required in processors implementing AMO and the RISC-V weak memory model, the hardware overhead for these instructions should be small.
\section{Experiment Methodology}
\label{sec:methodology}

This section provides an overview of the experimental setup for evaluating potential performance benefits of the \texttt{ld.aq} and \texttt{st.rl} instructions on workloads with a high-degree of sharing. Section \ref{ssec:gem5} describes how the gem5 architectural system simulator was altered in order to explore the performance impact of \texttt{fence} instructions with weaker ordering requirements on RISC-V executables. Section \ref{ssec:herd} describes how the RISC-V formal memory model was extended with exemplar programs (i.e., litmus tests) that demonstrate the new \texttt{ld.aq} and \texttt{st.rl} instructions ordering semantics by leveraging formal equivalence checking. Section \ref{ssec:clang} describes a headroom proxy study performed on an Apple Silicon AArch64 machine for evaluating potential performance gains on a modern lock-free concurrent queue implementation, which heavily utilizes the \texttt{std::memory\_order\_acquire} and \texttt{std::memory\_order\_release} constructs in order to achieve competitive performance under heavy-concurrency, single-producer-multi-consumer (SPMC), and multi-producer-single-consumer (MPSC) scenarios.

\subsection{Enhancing gem5 with RISC-V Extension}
\label{ssec:gem5}

The gem5 simulator~\cite{gem5_2020} is a cycle-accurate CPU simulator which has seen broad adoption in academia and industry. gem5 provides a generic out-of-order CPU model, o3, which is intended to be configurable for several different ISAs, including AArch64 and RISC-V.

RISC-V provides two opcodes, \texttt{custom-0} and \texttt{custom-1}, which are currently unused and guaranteed to not be used by future standard extensions. Therefore, we chose to use these opcodes to implement \texttt{ld.aq} and \texttt{st.rl} in the gem5 simulator. Since gem5 provides the \texttt{LDAR} and \texttt{STLR} instructions in AArch64, as well as the AQ and RL flags in the \texttt{AMO} RISC-V instruction, existing structures can be reused to implement these instructions.

\subsection{Adding Support in the RISC-V Formal Memory Model}
\label{ssec:herd}

The \texttt{herd}~\cite{alglaveHerdingCatsModelling} suite of tools implements a axiomatic, formal model of the RVWMO (as well as RVTSO) memory model. A formal model is necessary because it allows for defining precisely what behaviours are allowed and which are forbidden; a feature not available with the litmus tests provided for other architectures.\footnote{Although the the formal model currently differs slightly from the textual version of the specification, a bug which is currently being worked on.} The formal model also allows the automatic generation of litmus tests, as well as the creation of arbitrary multithreaded programs and evaluating candidate executions to see if they would be allowed by the memory model. It also allows proposed modifications of the memory model (such as adding \textit{load-acquire} and \textit{store-release} instructions) to be evaluated by seeing how different executions are allowed or forbidden by the adjustments.

Contrary to the claims on the mailing list~\cite{techUnprivilegedListsRiscv}, \texttt{herd} largely already supported \texttt{ld.aq} and \texttt{st.rl} in the parser for RISC-V, with the notion of release and acquire already present to support the annotations on AMOs (and on other architectures like AArch64 which already have \textit{load-acquire} and \textit{store-release}). However, there were no litmus tests included (for either base RISC-V or the \textit{load-acquire} or \textit{store-release} instructions). Therefore, we added tests to exercise these instructions and verify that they allowed the expected behaviour. 

\subsection{Atomics Performance Proxy Study on AArch64 Platform}
\label{ssec:clang}

The ideas behind \textit{load-acquire} and \textit{store-release} instructions are not new; these instructions are present on modern AArch64 platforms and available in consumer devices today. In order to understand how much performance can be gained by emitting \texttt{lw.aq} and \texttt{st.rl} instructions in scenarios where a full \texttt{fence} instruction is not required, a proxy study can be performed by modifying the clang C++ compiler to be conservative with memory ordering on existing AArch64 platforms. In other words, by modifying backend codegen for atomic instructions to enforce more ordering than required, program functional correctness is maintained, and the side-effects of additional ordering in the memory subsystem can be measured. To enforce additional ordering in the memory hierarchy, the target-independent atomic expansion pass and the AArch64 instruction selection pass are altered to bracket all atomic load/store operations with leading/trailing \texttt{fence} instructions. Since LLVM handles \texttt{\_\_atomic} operations with builtins, we can safely alter the emission of atomic instructions without recompiling \texttt{libc++}, \texttt{glibc}, or other internal C++ toolchain dependencies. The use of compiler builtins for primitives that are often implemented with target-specific intrinsics (including atomic operations) is common-practice, and enables target-aware optimizations to be applied without creating C++ toolchain fragmentation.

An open-source performance benchmark for lock-free queue data-structures~\cite{concurrentqueue_project} was selected as the target for the sensitivity analysis described above. Codegen quality is analysed in order to verify the lock-free queue benchmark provides interesting stimulus for quantifying the impact of additional memory ordering on runtime performance. Specifically, instruction opcode, number of atomic loads/stores, and atomic instruction frequency serves as the primary signals for determining if the benchmark is representative. The lock-free queue benchmark is compiled using latest \texttt{clang} as a reference and with a modified \texttt{clang} with more conservative memory-ordering constraints on an Apple Silicon 2021 MacBook Pro with the M1 Pro chip. It is important that a native system with a weak-memory model is used for collecting performance results. Since Total Store Ordered (TSO) and Sequentially Consistent (SC) systems have implicit \textit{load-acquire} and \textit{store-release} semantics by default, the additional \texttt{fence} instructions in the instruction stream would most-likely be elided after the \texttt{fence} instructions are decoded.

Apple Silicon systems also provide optional TSO support for compatibility with pre-existing x86 programs through binary translation~\cite{appleRosetta22019}. Care was taken to ensure that the Apple Silicon system did not have TSO implicitly enabled, which would negatively impact the performance results of the proxy study. An example project \texttt{is\_tso\_enabled} is provided in order to verify that TSO support has been properly disabled before benchmark results are captured.

The following section provides an in-depth discussion of the findings from attempting to improve Gem5 \textit{load-acquire} and \textit{store-release} handling, extending the RISC-V formal memory model to support explicit \textit{load-acquire} and \textit{store-release} instructions, and estimating the speedup of leveraging \texttt{ld.aq} and \texttt{st.rl} instructions on real programs.
\section{Results}
\label{sec:results}

Even though the scope of adding explicit \textit{load-acquire} and \textit{store-release} instructions appears well-defined, we encountered several obstacles that quickly increased the complexity of the original statement of work. This section is broken down into the three areas explored by this project. Section \ref{ssec:results_gem5} provides an overview of our findings from modifying the gem5 simulator to support the explicit \texttt{ld.aq} and \texttt{st.rl} instructions. Section \ref{ssec:results_mem_model} provides concrete examples of the ordering semantics that the \texttt{ld.aq} and \texttt{st.rl} instructions enforce. Section \ref{ssec:perf_results} provides an overview of the performance impact of the additional memory ordering on an AArch64 system with a weak memory model.

\subsection{Gem5 Support}
\label{ssec:results_gem5}

We added \texttt{ld.aq} and \texttt{st.rl} instructions to the RISC-V implementation in gem5 by overriding the \texttt{custom-0} and \texttt{custom-1} opcodes and reusing normal scalar load/store instruction semantics. However, simulator limitations mean that they are not handled in a usefully accurate way. As discussed previously, gem5 nominally supports acquire and release semantics for AArch64 and the RISC-V \texttt{AMO} instruction. However, analysis of the gem5 source code revealed that these operations are actually being treated as full barriers internally. There is no official documentation on how simulation of these operations is intended to work, but it appears that this is a known issue~\cite{gem5_la}.

From further analysis of the source code, it seems that there are unfinished efforts to add proper support for acquire and release operations to gem5, including \texttt{ACQUIRE} and \texttt{RELEASE} memory flags present in the ISA specifications, but these attributes are unused and no documentation exists on their intended function. Overall, gem5 seems to take the approach of sacrificing fine-grain modeling of memory ordering for ease-of-implementation and less complexity. In general, this seems like an acceptable tradeoff given that memory-ordering effects can be difficult to quantify, but it limits its usefulness for our analysis. While we did explore the possibility of adding proper support for acquire and release operations, it quickly became apparent that this would require major changes to the memory subsystem and be very difficult to verify. Although we did successfully generate RISC-V executables for our desired workloads, any performance data would be unreliable due to simulator limitations.

\subsection{Formal Memory Model Semantics}
\label{ssec:results_mem_model}

We created four litmus tests and associated diagrams to demonstrate, through the \texttt{herd} toolsuite.
They all revolve around a release-acquire between threads, with one thread releasing data and the other thread acquiring it.
\texttt{herd} has several operators to express relationships between events, the most important are \texttt{co} (coherence), representing the total order of all writes to a location, \texttt{rf} (reads-from), representing the propogation of data from a write to a read, \texttt{fr} (from-reads), representing a write that overwrites the data read by a read and hence must happen afterwards, and \texttt{ppo} (preserved program order), which represents the ISA semantics of what memory orderings must appear to execute in program order, for example through data dependencies or fences.

\begin{figure}[h]
 \includegraphics[width=\columnwidth]{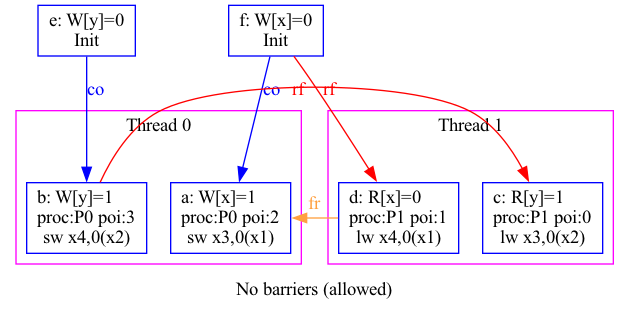}
 \caption{Allowed, although incorrect, execution}
 \label{fig:nobarriersfail}
\end{figure}

As an initial test of \texttt{herd}, we tried the program shown in Listing \ref{lst:nobarriers}.
This program deliberately has no barriers, and the incorrect execution, where the flag has changed but the data has not, is allowed and shown in Figure \ref{fig:nobarriersfail}.
The root of the problem is that there is no dependency between either the pair of loads or the pair of stores.

\begin{figure}[h]
 \includegraphics[width=\columnwidth]{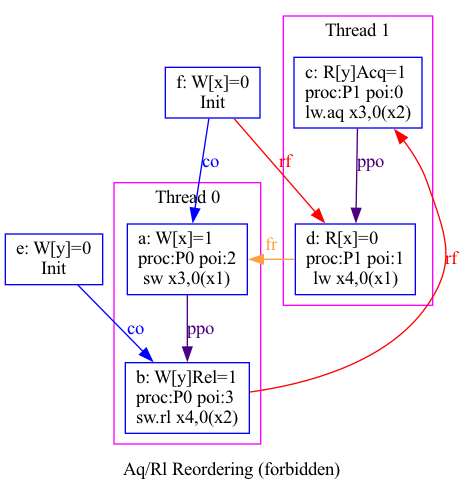}
 \caption{Incorrect execution is now forbidden}
 \label{fig:lwaqfail}
\end{figure}

Next, we ensured the \texttt{st.rl} and \texttt{ld.aq} instructions had the desired effect.
We added them in Listing \ref{lst:smartgood} to the appropriate instrutions, ending the sequence of stores with a \textit{store-release} and beginning the sequence of loads with a load-acquire.
That same execution is now disallowed as shown in Figure \ref{fig:lwaqfail}.
The incorrect execution is now disallowed because of the \texttt{ppo} (Preserved Program Order) edges between both the pair of stores and pair of loads.
Now, the incorrect execution implies a cycle which is disallowed by the memory model.

\begin{figure}[h]
 \includegraphics[width=\columnwidth]{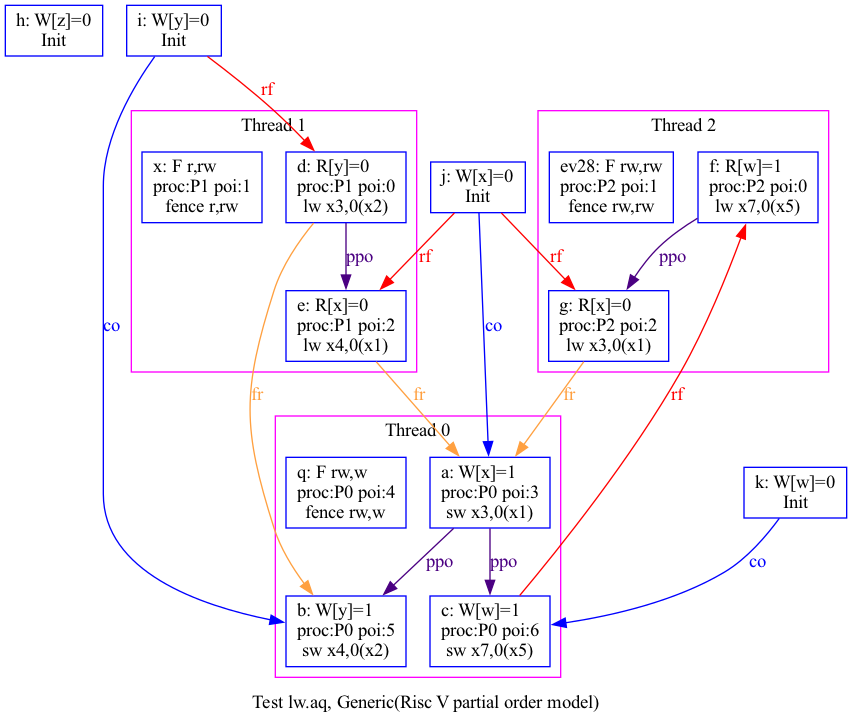}
 \caption{Forbidden, but desired, execution}
 \label{fig:falsecycleppo}
\end{figure}

Next, we explored the executions forbidden by the ``dumb'' implementation of \textit{load-acquire} and \textit{store-release} as normal loads and stores paired with fences.
The code is shown in Listing \ref{lst:dumbforbid}.
A third thread is used to attempt to observe moving the unrelated store to \texttt{w} ahead of the store to \texttt{x}.
As shown in Figure \ref{fig:falsecycleppo}, this (desired) reordering leads to a cycle because of the \texttt{ppo} edge between the writes to \texttt{w} and \texttt{x}.
That comes from the required \texttt{fence}, which enforces additional ordering beyond which is required in this scenario.

\begin{figure}[h]
 \includegraphics[width=\columnwidth]{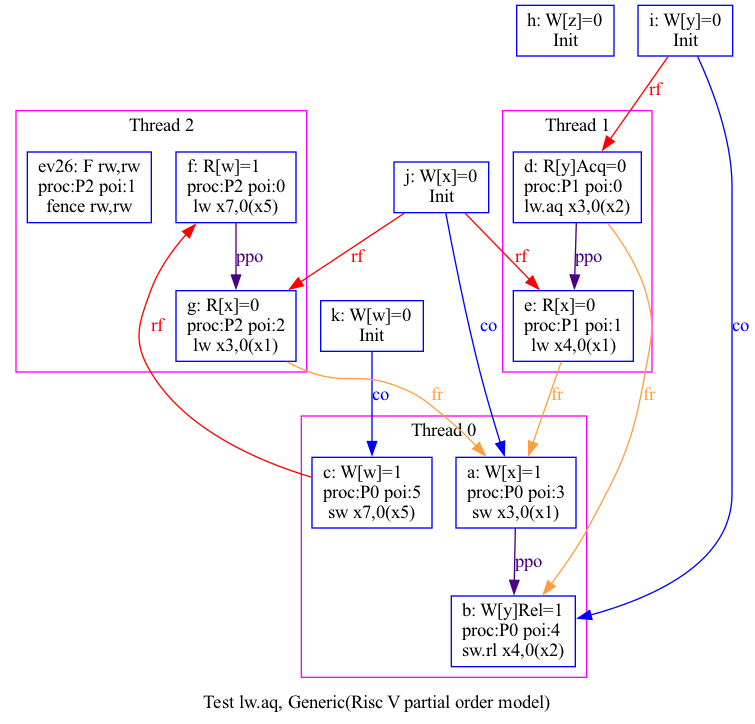}
 \caption{Allowed and desired execution}
 \label{fig:nofalsecycleppo}
\end{figure}

With the addition of \textit{load-acquire} and store-release, expressing only the required orderings is now possible.
Listing \ref{lst:smartallowed} shows the upgraded code. Note that this is not a safe transformation in general, since if the ordering between \texttt{w} and \texttt{x} \textit{was} desired, then it would be lost.
Now, as seen in Figure \ref{fig:nofalsecycleppo}, there is no \texttt{ppo} between the stores and this execution is legal.

\subsection{Performance Benchmark Results}
\label{ssec:perf_results}

\begin{figure*}[ht]
 \includegraphics[width=\textwidth]{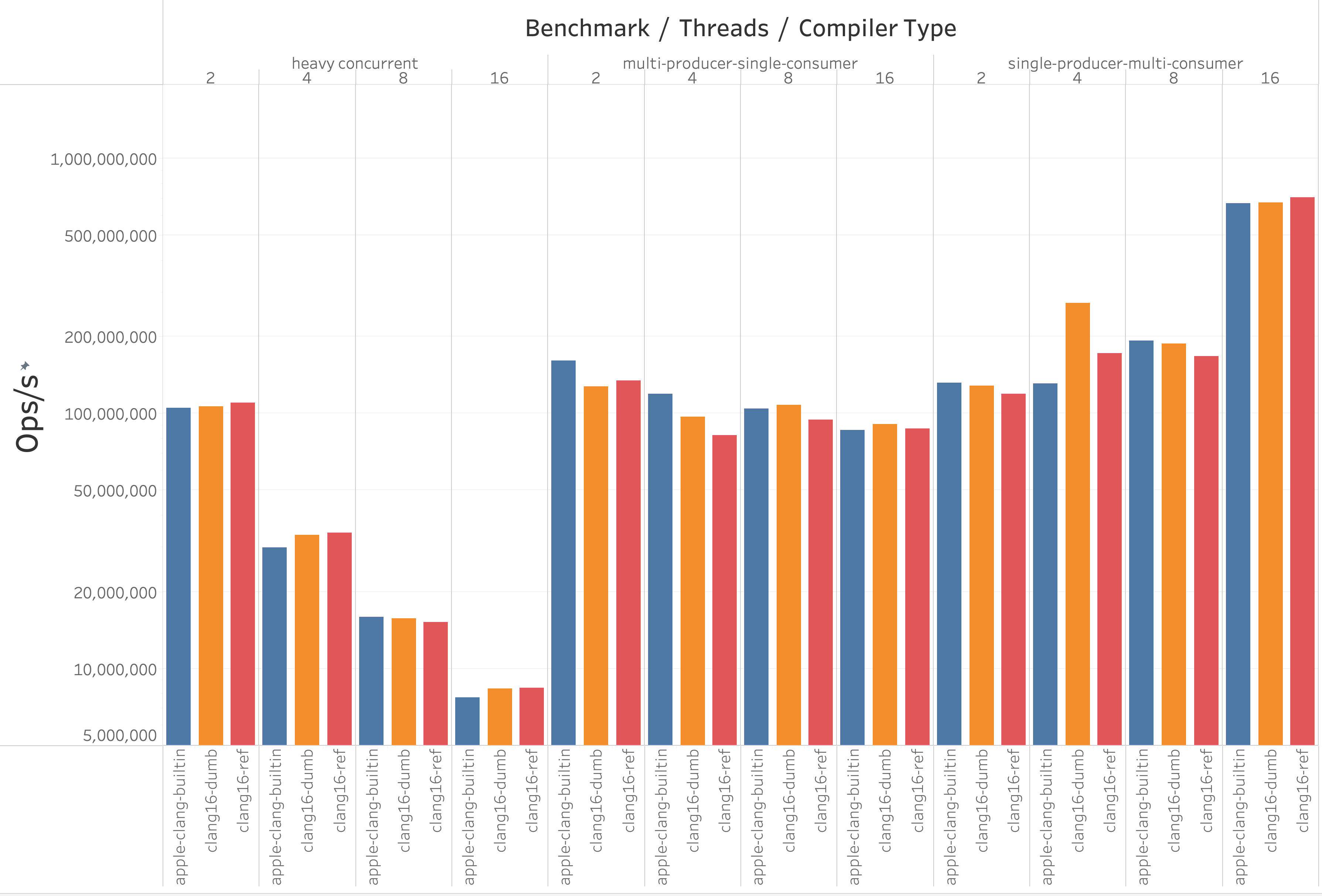}
 \caption{Total Queue Operations Per Second by Compiler Type on AArch64.}
 \label{fig:ops_per_second_by_compiler}
\end{figure*}

\begin{figure}[h]
 \includegraphics[width=\columnwidth]{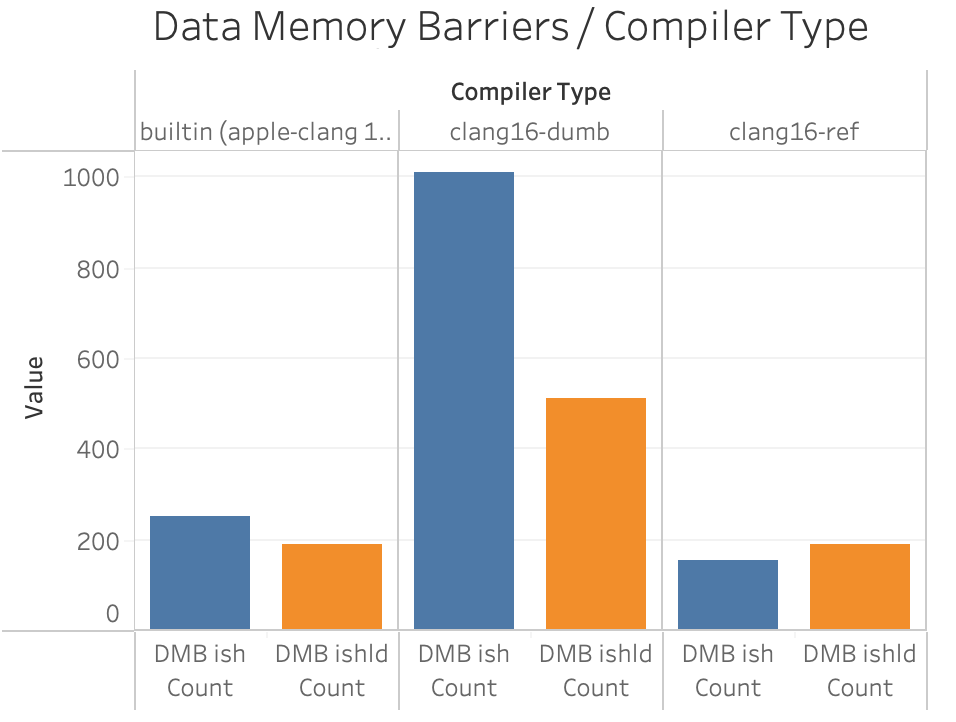}
 \caption{Total memory barriers emitted by each codegen strategy.}
 \label{fig:dmbs_by_compiler_type}
\end{figure}

In this section, the results from the memory-ordering performance proxy study on an Apple AArch64 platform (M1 Macbook Pro) are summarized. Figure \ref{fig:ops_per_second_by_compiler} shows the total number of queuing operations (i.e., both \texttt{enqueue} and \texttt{dequeue}) per second performed by each of the different compiler backends in the study. Because \textit{enqueue} and \textit{dequeue} operations require ordering semantics for updating head/tail pointers, the number of queueing operations per second loosely translates to the number of atomic operations performed by the selected workload. Unexpectedly, getting rid of \textit{load-acquire} and \textit{store-release} instructions and instead forcing the compiler to emit only bidirectional \texttt{fence} instructions improved performance in some scenarios. In order to verify that our compiler modifications for more conservative memory were correctly impacting downstream instruction emission, the disassembly can be analyzed in order to ensure that additional memory barrier instructions are being emitted by the compiler-under-test. Figure \ref{fig:dmbs_by_compiler_type} shows that our \texttt{clang-dumb} implementation with conservative memory ordering emits $\approx4.5$ times as many data barriers as the reference clang compiler. The large increase in total barriers emitted by the compiler is a strong signal that our modifications to the backend were successful, and that we are enforcing more ordering in the memory hierarchy. The \texttt{dmb ish} instruction is a data memory barrier (inner shareable domain), it is a full fence that only applies to the CPU cores. The \texttt{dmb ishld} is a data memory barrier (inner shareable domain, load) is a load-only fence. 

We believe that this is because of our choice of benchmark.
Consulting with experts in the field has led us to the conclusion that in this benchmark, with a large amount of data sharing between threads and hence cores, the process of actually moving the data through the memory systems dominates and is unable to be covered by the processor.
Therefore, the additional flexibility that is available to the memory system cannot be used, and so we are only seeing interference patterns and noise in the data.
Benchmarks with less actual data movement, such as lightly-contented locks, may have served as a better way to contrast between the methods.
Additionally, we were surprised to find that upstream \texttt{clang-16} compiler outperformed the native Apple \texttt{clang} proprietary toolchain. 

\section{Future Work}
\label{sec:future_work}

As the upstream RISC-V community is already pushing to ratify the explicit \texttt{ld.aq} and \texttt{st.rl} instructions, future work for this project is to ensure that these instructions are ratified and exposed in the RISC-V Platform Specification~\cite{riscvPlatformSpecs2022}. Once the \textit{load-acquire} and \textit{store-release} instructions are ratified and added to the ABI requirements defined by the platform spec, work will begin on adding robust support for them in compiler backends. Other future work includes enhancing the gem5 simulator to support finer granularity modeling of acquire and release semantics in weak memory models, integrating our memory model litmus tests into the upstream \texttt{herd} tool suite, and collecting performance results on native RISC-V silicon hardware.
\section{Conclusion}
\label{sec:conclusion}
In this project, we explored adding load-acquire and store-release operations to the RISC-V ISA. RISC-V provides a Weak Memory Ordering model (RVWMO) which enables simplification of the memory subsystem but places the burden of explicit software synchronization on the programmer. Load-acquire and store-release are an alternative to traditional (stronger) software memory barriers which expose additional opportunities for parallelism in the memory hierarchy subsystem. First, we explored adding support in the gem5 simulator for explicit load-acquire and store-release instructions by extending the RISC-V ISA. Although we correctly introduced new instructions to the ISA, we were unable to make a meaningful comparison due to simulator limitations - gem5 is currently incapable of representing load-atomic and store-release instructions at the level of fidelity needed to observe performance characteristics.
Second, we enhanced the \texttt{herd} formal memory model toolchain by encoding the memory ordering constraints that the explicit load-acquire and store-release instructions enforce. Although much of what was needed was already present to support atomic \texttt{AMO} operations, we extended the model with explicit support and added four new ``litmus tests'' intended to demonstrate the new instruction semantics. We found that framing memory ordering behavior as a formal equivalence check against known specifications is a powerful technique for testing the correctness of weak memory models.
Finally, we performed a proxy study on Arm silicon by developing a ``dumb'' version of the LLVM/Clang toolchain which was incapable of producing load-acquire and store-release instructions, and evaluated performance (measured in atomic operations per second) against upstream Clang and Apple's own proprietary toolchain. We found that in many cases using full barriers actually \textit{improved} performance, suggesting that compiler support for these operations is immature.

%
%
%
%
%
%
%
%
%
%
%

\bibliographystyle{IEEEtran}
\bibliography{main.bib}

\clearpage
\appendix
Litmus test specifications for the \texttt{herd} formal model:
\begin{minipage}{\columnwidth}
\begin{lstlisting}[caption=No barriers,label={lst:nobarriers}]
RISCV no barrier fail
{
0:x1=x; 0:x2=y;
1:x1=x; 1:x2=y;
x=0; y=0;
}
P0                 | P1              ;
addi x3, x0, 1     | lw    x3, 0(x2) ;
addi x4, x0, 1     | lw    x4, 0(x1) ;
sw   x3, 0(x1)     |                 ;
sw   x4, 0(x2)     |                 ;
exists (1:x3=1 /\ 1:x4=0)
\end{lstlisting}
\end{minipage}

\begin{minipage}{\columnwidth}
\begin{lstlisting}[caption=Correct behaviour with acquire/release,label={lst:smartgood}]
RISCV lw.aq sw.rl
{
0:x1=x; 0:x2=y;
1:x1=x; 1:x2=y;
x=0; y=0;
}
P0                 | P1              ;
addi x3, x0, 1     | lw.aq x3, 0(x2) ;
addi x4, x0, 1     | lw    x4, 0(x1) ;
sw    x3, 0(x1)    |                 ;
sw.rl x4, 0(x2)    |                 ;
exists (1:x3=1 /\ 1:x4=0)
\end{lstlisting}
\end{minipage}

\begin{minipage}{\columnwidth}
\begin{lstlisting}[caption=Desired but disallowed behaviour,label={lst:dumbforbid}]
RISCV lw.aq sw.rl dumb not allowed
{
0:x1=x; 0:x2=y; 0:x5=w; 0:x6=z;
1:x1=x; 1:x2=y; 1:x5=w; 1:x6=z;
2:x1=x; 2:x2=y; 2:x5=w; 2:x6=z;
x=0; y=0; w=0; z=0;
}
P0              | P1               | P2 ;
addi x3, x0, 1  | lw    x3, 0(x2)  |    ;
addi x7, x0, 1  | fence r,rw       |    ;
addi x4, x0, 1  | lw    x4, 0(x1)  |    ;
sw    x3, 0(x1) |                |      ;
fence rw,w      |              |        ;
sw    x4, 0(x2) |            |          ;
sw    x7, 0(x5) |          |            ;
                |        | lw x7, 0(x5) ;
                |        | fence rw, rw ;
                |        | lw x3, 0(x1) ;
exists (2:x3=0 /\ 2:x7=1)
\end{lstlisting}
\end{minipage}

\begin{minipage}{\columnwidth}
\begin{lstlisting}[caption=Allowed with acquire/release,label={lst:smartallowed}]
RISCV lw.aq sw.rl smart allowed
{
0:x1=x; 0:x2=y; 0:x5=w; 0:x6=z;
1:x1=x; 1:x2=y; 1:x5=w; 1:x6=z;
2:x1=x; 2:x2=y; 2:x5=w; 2:x6=z;
x=0; y=0; w=0; z=0;
}
P0              | P1               | P2 ;
addi x3, x0, 1  | lw.aq x3, 0(x2)  |    ;
addi x7, x0, 1  |                  |    ;
addi x4, x0, 1  | lw    x4, 0(x1)  |    ;
sw    x3, 0(x1) |                |      ;
                |              |        ;
sw.rl x4, 0(x2) |            |          ;
sw    x7, 0(x5) |          |            ;
                |        | lw x7, 0(x5) ;
                |        | fence rw, rw ;
                |        | lw x3, 0(x1) ;
exists (2:x3=0 /\ 2:x7=1)
\end{lstlisting}
\end{minipage}

\end{document}